# DNA origami route for nanophotonics

Anton Kuzyk[1,2], Ralf Jungmann[3,4], Guillermo P. Acuna[5], and Na Liu[1,6*]

[1]Max Planck Institute for Intelligent Systems, Heisenbergstrasse 3, D-70569 Stuttgart, Germany
[2]Department of Neuroscience and Biomedical Engineering, Aalto University School of Science, P.O. Box 12200, FI-00076 Aalto, Finland
[3]Department of Physics and Center for Nanoscience, Ludwig Maximilian University, 80539 Munich, Germany
[4]Max Planck Institute of Biochemistry, 82152 Martinsried near Munich, Germany
[5]Institute for Physical & Theoretical Chemistry, and Braunschweig Integrated Centre of Systems Biology (BRICS), and Laboratory for Emerging Nanometrology (LENA), Braunschweig University of Technology, 38106 Braunschweig, Germany
[6]Kirchhoff Institute for Physics, University of Heidelberg, Im Neuenheimer Feld 227, D-69120 Heidelberg, Germany

*email: naliu@is.mpg.de

## ABSTRACT:

Specificity and simplicity of the Watson-Crick base pair interactions make DNA one of the most versatile construction materials for creating nanoscale structures and devices. Among several DNA-based approaches, the DNA origami technique excels in programmable self-assembly of complex, arbitrary shaped structures with dimensions of hundreds of nanometers. Importantly, DNA origami can be used as templates for assembly of functional nanoscale components into three-dimensional structures with high precision and controlled stoichiometry. This is often beyond the reach of other nanofabrication techniques. In this Perspective, we highlight the capability of the DNA origami technique for realization of novel nanophotonic systems. First, we introduce the basic principles of designing and fabrication of DNA origami structures. Subsequently, we review recent advances of the DNA origami applications in nanoplasmonics, single-molecule and super-resolution fluorescent imaging, as well as hybrid photonic systems. We conclude by outlining future prospects of the DNA origami technique for advanced nanophotonic systems with tailored functionalities.

# Introduction

Precise arrangement of individual photonic building blocks in space, including metal and semiconducting nanoparticles (NPs), quantum dots (QDs), nanodiamonds, fluorophores, etc., is crucial for creation of advanced nanophotonic systems with tailored optical properties and novel functionalities. Despite the remarkable advances in nanophotonics enabled by top-down fabrication techniques, critical limitations still remain, for instance, realization of three-dimensional complex nanostructures, especially with structural reconfiguration as well as organization of nanoscale components of different species in close proximity.

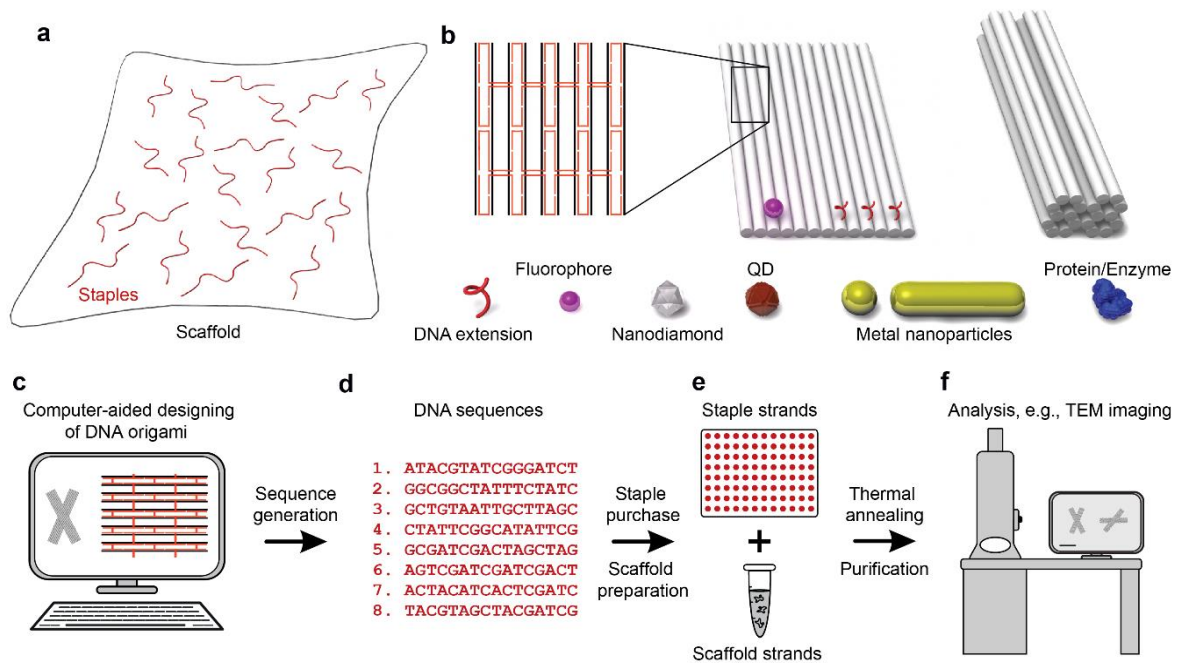

**Figure 1 | Principle of the DNA origami folding, design, and assembly. (a)** DNA origami consists of long single-strand DNA ('scaffold'), and several hundreds of short ssDNA strands ('staples'). **(b)** Upon thermal annealing, the 'staples' fold the 'scaffold' strands into two- or three- dimensional structures with predesigned shapes. DNA origami structures can be modified with ssDNA extensions that serve as binding sites for further assembly of different nanoscale components, including fluorophores, quantum dots, nanodiamonds, metal nanoparticles, proteins, etc, into almost arbitrary geometries with nanometer precision. **(c)** DNA origami structures are usually designed with caDNAno software. **(d)** Sequences of the 'staple' strands required for the assembly are generated. **(e)** 'Staple' strands are usually purchased in multiwall plates from commercial vendors specializing in automated DNA synthesis. Single-stranded phage DNA is typically used as scaffold for DNA origami structures. The scaffold strands can be produced by M13 phage amplification or purchased from several suppliers. Scaffold is mixed with 'staple' strands (with a large excess) and the origami structures are assembled through thermal annealing. The structures are usually purified before being used as templates for further assembly. **(f)** Atomic force microscopy (AFM) and transmission electron microscopy (TEM) are often used to characterize two- and three-dimensional origami structures.

Molecular self-assembly offers an alternative to circumvent these limitations[1,2]. In particular, the DNA origami technique[3–6] identifies a unique route for realization of nanophotonic structures with hierarchical complexities. DNA origami can be created in almost any arbitrary shapes. Such origami structures can then serve as templates for assembly of a variety of functional components[7] with nanoscale precision.

Figure 1 illustrates the workflow of the DNA origami formation. Long single-stranded DNA (ssDNA) with known sequence (called 'scaffold' and derived from the single-stranded genome of the M13 bacteriophage) is mixed with a set of short synthetic ssDNA (called 'staples' and usually purchased from oligonucleotide synthesis vendors) (Figure 1a). Each 'staple' strand possesses a unique sequence and binds the scaffold at specific positions. This makes DNA origami fully addressable. The staple strands fold the scaffold strand into a predesigned two- or three-dimensional shape (Figure 1b). ssDNA extensions (called capture strands) or chemical modifications, e.g., biotin and amino groups on origami can serve as binding sites for precise arrangement of nanoscale components including proteins[8–13], NPs[14–22], fluorophores[23–27], which are functionalized with complementary binding modifications (Figure 1b). Fluorophores can also be directly incorporated into the DNA 'staple' strands.

Generally, the workflow of a DNA origami-templated nanophotonic system starts with identification of individual components of interest and their desired relative spatial arrangement. Subsequently, a DNA origami structure with certain geometry is conceived to template such an arrangement. The DNA origami structure is designed using computer-aided design (CAD) software, e.g., caDNAno[28] (Figure 1c). As output, a set of DNA sequences for the staple strands (Figure 1d) are generated and sent to commercial vendors specializing in automated DNA synthesis (Figure 1e). The obtained staple strands are mixed with the scaffold strand of choice (produced by M13 phage amplification[29,30] or purchased) followed by thermal annealing. The origami structures are then purified[31–35] for structural characterizations (Figure 1f). Functionalization of individual components that can bind to the origami often utilizes conjugation with ssDNA strands[36–39], complementary to the capture strands. Another frequently used method is incorporation of biotin/streptavidin modifications that can direct bind to the origami through streptavidin-biotin interactions[16,20,40,41]. Less often used are alkyne, amino and azido functional groups[42,43].

There are excellent reviews and perspectives that elucidate the technical aspects of the DNA origami technique, including designing, assembly, and characterization of DNA origami structures[31,34,44]. Since its birth, the DNA origami technique has been widely used in a wealth of research fields[6,7,45,46] for instance, drug delivery[47], artificial nanopores[48–50], single molecule studies[51,52], macroscopy standards[27,53,54], etc. In this perspective, we will focus on the applications of DNA origami in nanophotonics. Particularly, we highlight several promising directions, along which the DNA origami technique may help to solve the present technological challenges and opens new pathways to realizing nanophotonic systems with novel functionalities.

## DNA origami for nanoplasmonics

Localized surface plasmon resonances result from collective oscillations of the conduction electrons in metal NPs, when they interact with light. The plasmon resonances can be tuned by the compositions, shapes, and local surroundings of the metal NPs. Plasmons of metal NPs placed in close proximity can be coupled, mixed, and hybridized[55]. Such coupling is very sensitive to the relative arrangement of the individual NPs in space. On one hand, this provides a unique opportunity to engineering near- and far-field optical properties of the constructed nanostructures[56–58]. On the other hand, it poses many technical challenges to assemble metal NPs into well-defined configurations, especially in three dimensions[59].

At the end of the 20th century, DNA emerged as one of the most versatile construction materials at the nanoscale[60]. Utilization of DNA for assembly of metal NPs into larger structures was first demonstrated by Alivisatos and Mirkin in 1996[61,62]. Since then, DNA has been widely used for direct assembly of NPs into a variety of structures with increasing complexities[18,63–66]. Impressive progresses have been witnessed in DNA-based assembly of two- and three-dimensional periodic lattices[67–73]. However, fabrication of well-defined plasmonic clusters composed of discrete numbers of interacting metal NPs remained challenging until the introduction of the DNA origami technique in 2006 by Rothemund[3]. Inherent addressability of the DNA origami made it ideally suitable for templated assembly of plasmonic nanostructures. Nevertheless, several technical challenges had to be overcome in order to achieve plasmonic systems with distinct optical properties. The first advancement was realization of high-yield assembly of NPs on DNA origami templates[14,74–76] (Figure 2a-d). Initially, single-layer DNA origami and spherical gold nanoparticles (AuNPs) were widely used due to the ease of design and fabrication. DNA conjugation with AuNPs was done through gold-thiol bond[15,74,77]. Soon after this, methods for assembly of silver spherical NPs[76] (Figure 2d) and anisotropic gold nanorods (NRs) were developed[78]. In addition, further advancement of the origami technique, for instance, extension into three dimensions[4] and introduction of twisted and curved structures[79], enabled fabrication of DNA origami-templated assemblies of metal NPs[20,80–82] with unprecedented complexities (Figures 2e-h).

Apart from the well-established approach to assemble metal NPs on DNA origami with ssDNA capture strands[83], there are also several other solutions. Attempts were made to metalize the entire origami structures[84,85] (Figure 2i). For metallization, DNA templates were first seeded with small gold or silver clusters followed by electroless deposition of gold for further metal growth. Electroless metal deposition not only enlarged the size of the NPs [86] but also could fuse the particles together[87] (Figure 2j). DNA origami structures were also used as molds for growth of metal colloids with complex shapes[88,89]. A small particle served as seed (Figure 2k) and the origami structure restricted the growth of the metal into a specific and predefined shape (Figure 2l). In addition, standard silicon etching and metal deposition techniques were used to transfer the shape of two-dimensional DNA origami objects into metal nanostructures[90].

The ability to arrange metal NPs on DNA origami templates with high yield and accuracy opened the pathway to construct plasmonic structures with novel optical properties. In 2012, Liedl and Ding used DNA origami to arrange metal NPs into helical assemblies[86,91], respectively, (see Figure 2m) which exhibited strong plasmonic circular dichroism in the visible spectral range, originated from collective interactions of the metal NPs in the helical geometry[92]. Since then, DNA origami has been widely used as templates for assembly of spherical NPs[81,82,93,94] and NRs [95–98] (Figure 2n-p) into chiral plasmonic structures[99–101]. Utilization of NRs provides an additional benefit for generating stronger optical response compared to spherical NPs[102]. However, assembly of NRs on DNA origami is technically more challenging. Another example of origami-based plasmonic nanostructures with novel optical responses is a ring of NPs, which exhibited both electric and magnetic resonances at visible frequencies[80] (Figure 2f). Other progresses have been also achieved in fabrication of NP-based waveguides for energy transfer[103–105].

Importantly, DNA origami-templated plasmonic nanostructures is not limited to static systems[106]. Solution-based nature of DNA structures and dynamic DNA nanotechnology[107,108] provide a unique way to actively manipulate both spatial and temporal arrangements of metal NPs, enabling reconfigurable plasmonic systems with dynamically controlled optical responses. Liedl group demonstrated reversible plasmonic circular dichroism responses by orientation switching of the origami-templated chiral plasmonic assemblies on surface[109]. Reconfigurable plasmonic structures were also realized by assembly of metal NPs on dynamic DNA origami templates, which were switched among several configurations by external stimuli (Figure 2o). The spatial reconfiguration of the DNA origami templates resulted in rearrangement of the plasmonic NPs and therefore altered optical responses[110–113]. There are various ways to control the configurations of the DNA origami templates[106]. Probably, the most versatile and thus widespread approach is based on the so-called "toehold-mediated strand displacement reaction"[107], which utilizes DNA strands as fuel to regulate spatial configurations[110]. Also, photoresponsive molecules such as azobenzene can be employed through incorporation with DNA to activate responses upon light stimuli[111,114]. Recently, selective manipulation of DNA origami-based plasmonic structures has been demonstrated, taking the advantage of the pH sensitivity of the DNA triplexes[115,116]. More intriguing approaches could include reversible reconfiguration based on shape-complementarity[117,118] or structural adaptions of aptamers to the presence of target molecules[119–121]. In addition to using reconfigurable DNA origami templates, it is also possible to rearrange metal NPs on static DNA origami templates (see Figure 2p). In this approach, NPs are either guided by the so-called molecular walkers[122] or act as walkers themselves[123,124].

**Figure 2 | DNA origami for nanoplasmonics. (a-d)** Examples of metal NP assemblies templated by DNA origami with high yield and accuracy[74–77]. **(e-h)** Sophisticated metal NP assemblies on three-dimensional complex

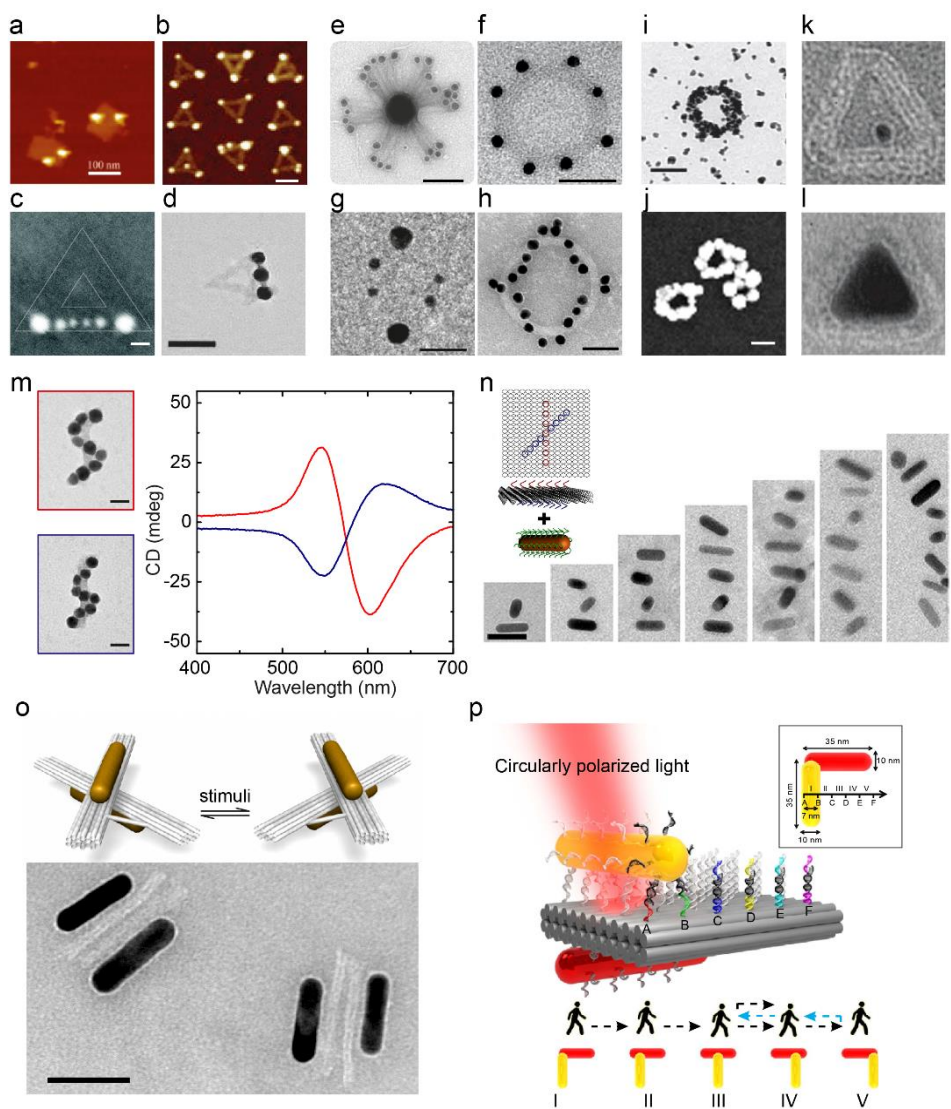

origami templates [20,80–82]. **(i)** Metallization of the DNA origami ring structures[84]. **(j)** Metal NPs fused together on DNA origami through electroless metal deposition[87]. **(k,l)** DNA origami structures as molds for growth of metal colloids with defined morphologies[88]. **(m)** DNA origami-templated assembly of helical NP assemblies with strong plasmonic circular dichroism in the visible spectral range[86]. **(n)** Chiral plasmonic assemblies with gold NRs[98]. **(o)** Plasmonic nanostructures with dynamically controlled optical responses enabled by stimulus-driven DNA origami templates[110,111,115]. **(p)** Plasmonic walker on DNA origami[123]. Scale bars: **(a-c)** 100 nm, **(d)** 20 nm, **(e)** 100 nm, **(f)** 50nm, **(g)** 30 nm, **(h)** 50nm, **(i,j)** 100nm, **(m)** 20 nm, **(n,o)** 50 nm.

# DNA origami for fluorescence imaging

Since its invention, DNA origami has found numerous applications in the field of fluorescence imaging[125], owing to its bottom-up self-assembly properties and the availability of dye-modified oligonucleotides. Especially, the combination of DNA origami nanostructures with single-molecule fluorescence techniques is attractive due to origami's unique spatial addressability on length scales ranging from a few to hundreds of nanometers combined with exquisite positioning accuracy. One of the first applications of DNA origami in single-molecule fluorescence was its combination with emerging super-resolution techniques[126–128] to create nanoscopic rulers for resolution calibration[129]. Stochastic super-resolution techniques circumvent the classical diffraction limit of light by "switching" fluorophores from the so-called dark- to bright-states and back, thus only activating and localizing the emission of a single dye molecule in a diffraction-limited area at every given point in time. Time-lapsed acquisition and repeated switching then eventually allows for complete reconstruction of all molecule positions, yielding a super-resolution image. While super-resolution techniques readily achieve spatial resolutions down to a few tens of nanometers, it is hard to precisely quantify their achievable spatial resolution due to the lack of versatile nanoscale rulers. However, DNA origami nanostructures are ideal calibration standards[53,129] due to their high folding yield and sub-nanometer positioning accuracy.

DNA origami nanostructures are also ideally suited as test structures for the development of new imaging approaches. While stochastic super-resolution techniques such as Stochastic Reconstruction Microscopy (STORM)[128] or Photoactivated Localization Microscopy (PALM)[127] are already starting to transform the way we look at biology today, their experimental implementation – especially with regards to multiplexed detection, i.e. the imaging of multiple targets – is still challenging due to the necessity to carefully adjust buffer conditions for each fluorophore species. DNA Points Accumulation In Nanoscale Topography (DNA-PAINT)[130–140] was developed to overcome some of the difficulties of incumbent super-resolution approaches. In DNA-PAINT, stochastic "blinking" of targets is achieved by the transient hybridization of short, dye-labeled oligonucleotides (called "imager" strands) to their complementary strands ("docking" strands) on a target of interest (Figure 3a). Unbound imager strands freely diffuse in solution, adding only minimal background when image acquisition is performed in total internal reflection or oblique illumination[141]. As DNA-PAINT uses transient hybridization of short oligonucleotides to create the necessary blinking in stochastic reconstruction microscopy, it is ideally suited to visualize DNA nanostructures. With obtainable spatial resolution on the nanometer scale, features such as the two faces of the DNA origami structure – spaced only 16 nm apart – are clearly resolvable (Figure 3b).

In DNA-PAINT, imaging and labeling probes can actually be seen as DNA barcodes owing to their unique sequences of the DNA bases. Thus, multiplexing can be easily achieved by sequential imaging of different target molecules labeled with orthogonal docking strands (Figure 3c). In this approach, called Exchange-PAINT[137], the first target (e.g. P1) is imaged by the complementary strand (e.g. P1*, Figure 3c). Then a washing buffer is introduced to remove P1* from the sample, followed by the introduction of P2* imager strands to visualize the second target. This imaging and washing procedure is repeated until images for all targets are successfully acquired and pseudo-colors were assigned. Exchange-PAINT now enables spectrally-unlimited multiplexing, only restricted by the amount of orthogonal DNA sequences, which could easily reach hundreds under appropriate experimental conditions. Similar exchange strategies can be applied for other super-resolution approaches as well, using slightly more stable hybridization probes in combination with mild denaturation during washing rounds[142].

Due to DNA-PAINT's resistance to photobleaching (imager strands are constantly replenished from solution), very high spatial resolutions are achievable by extracting the maximum number of photons from a dye-labeled strand before unbinding from its target. In combination with intricate drift correction[139], DNA-PAINT achieves molecular-scale spatial resolutions of better than 5 nm, as demonstrated by imaging the MPI and LMU logo on DNA nanostructures shown in Figure 3d with single binding sites spaced 5 nm apart[143]. 3D super-resolution imaging of complex 3D DNA origami nanostructures is also straightforward to implement by using a cylindrical lens in the microscope imaging path to encode 3D location in an elliptical point spread function[136] (Figure 3e).

Further applications of DNA-based super-resolution imaging have demonstrated quantitative target detection *in vitro* and *in situ* in single cells, allowing researchers to count integer numbers of biomolecules based on their kinetic signature without spatially resolving them[140], even allowing single nucleotide mismatch discrimination of RNA targets[138]. Also, DNA origami is ideally suited for applications that do not require super-resolution. In a recent study, DNA origami structures have been used to enable the construction of novel fluorescent probes, termed "metafluorophores", that enable diffraction-limited imaging with up to 124 distinct colors[144]. This was achieved by using the exquisite spatial arrangement accuracy of origami to prepare objects with defined number of dyes, thus allowing for the construction of intensity barcodes (Figure 3f).

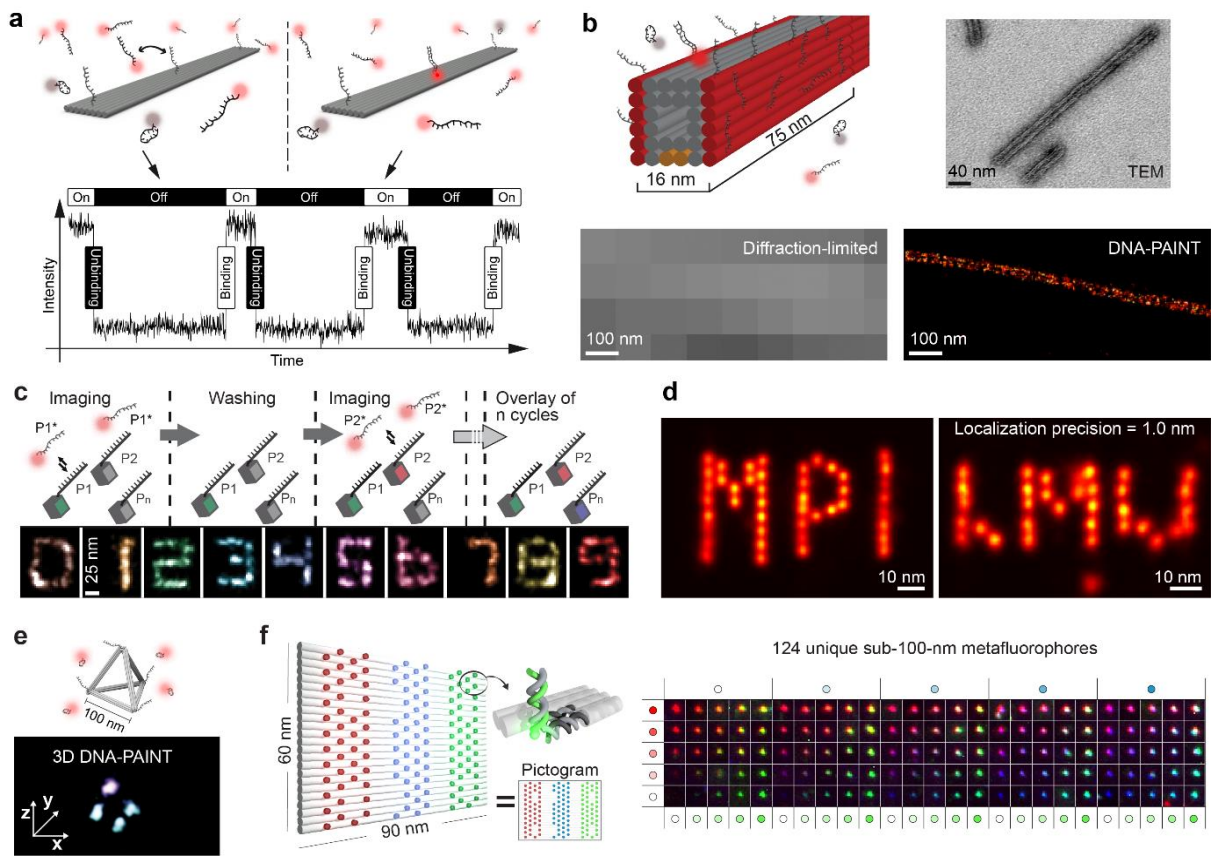

**Figure 3 | DNA origami for fluorescence applications. (a)** Super-resolution imaging by DNA-PAINT. Short dye-labeled oligonucleotides bind transiently to their complementary target on a DNA origami structure. The transient binding creates an apparent "blinking" used for stochastic super-resolution microscopy[130]. **(b)** Tunnel-like DNA origami with DNA-PAINT docking strands at red-colored faces imaged using transmission electron microscopy, diffraction-limited microscopy, and DNA-PAINT super-resolution microscopy[137]. **(c)** Exchange-PAINT enables spectrally-unlimited super-resolution multiplexing. Distinct molecular targets are labeled with orthogonal docking strands. Subsequent imaging is performed sequentially with alternating washing and imaging rounds using the same color fluorophore for all rounds[137]. **(d)** DNA-PAINT's resistance to photobleaching in combination with intricate drift correction enables ultra-resolution imaging, resolving docking strands spaced 5 nm apart on DNA origami structures[143]. **(e)** 3D-DNA-PAINT image of a DNA tetrahedron[136]. **(f)** Origami's precise control over spatial positioning and stoichiometry enables the construction of 124 "metafluorophores" for diffraction-limited barcoding applications using only three spectral colors and five prescribed intensity levels[144].

# DNA origami for hybrid photonic structures

One of the main advantages of the DNA origami technique lies in its capacity to self-assemble different species with precise stoichiometry control and nanometer precision. Perhaps one of the simplest examples of a hybrid photonic structure consists of a pair of fluorophores placed in close proximity. For distances typically below 10 nm, fluorescence resonance energy transfer (FRET) can occur between the fluorophores. The incorporation of fluorophores to DNA origami structures is straightforward. Staple strands labelled at a desired position with a palette of fluorophores across the visible range are commercially available. Most fluorophores are incorporated to the DNA sequences forming the staple strands through one linker. Although this approach guarantees a nanometer positioning of the fluorophore within the DNA origami structure, its orientation cannot be determined and will depend among other factors on the resulting fluorophore's charge.

Figure 4a depicts a fluorophore system self-assembled onto rectangular DNA origami. The precise arrangement of the fluorophores enables light guiding *via* FRET[145] with an energy path that can be controlled by the presence of a fluorophore acting as a "jumper". Furthermore, the DNA origami technique has also been exploited to self-assemble fluorophore systems for fabrication of artificial light harvesting antennas[23,25].

In addition to combination of fluorophores, another type of widespread hybrid photonic structures comprises optical light sources coupled to optical antennas[146]. Typical examples of the light sources include fluorophores and QDs, whereas optical antennas generally consist of nanometer-sized metal structures that exhibit localized surface plasmon resonances in the optical range. It is worth discussing the advantages of the DNA origami technique for the fabrication of this type of structures. Optical antennas have been successfully fabricated using ion or electron beam lithographic techniques[147]. However, these top-down approaches have several shortcomings. First, fabrication is serial by nature and requires specialized and costly equipment. Second, they often yield rough surfaces and polycrystalline metals with grains, which hamper the properties of the designed structure and reduce the resonance quality [148]. However, the most critical limitation of these techniques is that they are extremely challenging to position a single light source at the focus of an optical antenna[149]. In one example, a demanding multi-step lithographic procedure was employed to place single QD at the focus of a Yagi-Uda antenna[150], whereas in another work an undefined number of fluorophores were immobilized with the aid of a polymer layer in a region including the focus of a bow-tie antenna[147]. In contrast to the top-down approaches, the bottom-up DNA origami technique can overcome the aforementioned shortcomings. It is parallel in nature and capable of self-assembling colloidal crystalline metal NPs with higher quality resonances. Finally, both single light sources and optical antennas can be self-assembled with nanometer precision and stoichiometric control.

The first experiments in this direction are sketched in Figure 4b. The same rectangular DNA origami structure was used to study the distance dependent energy transfer between a single fluorophore and a single 10 nm AuNP[151] (Figure 4c). This approach enabled a detailed study of the manipulation of fluorescence with plasmonic NPs at the single molecule level, including the polarization in near field excitation[152], the controlled increase in photostability[153] together with the determination of how fluorescence rates are affected in the vicinity of NPs[154]. The flexibility of the DNA origami technique was also exploited to self-assemble dimer antennas (Figure 4d). Initial efforts were conducted towards fluorescence enhancement applications and included two AuNPs (with sizes up to 100 nm) and a single fluorophore at the hotspot self-assembled onto a 3D pillar shaped DNA origami structure[155]. The resulting gap between the NPs was higher than 20 nm which limited the fluorescence enhancement to two orders of magnitude. Additional developments on the NPs incorporation and origami design[156] lead to a reduction of the gap to approximately 10 nm and an increase of the fluorescence enhancement over three orders of magnitude[157]. Furthermore, with these results self-assembled optical antennas managed to outperform top-down lithographic antennas in terms of fluorescence enhancement and single molecule detection at elevated concentrations. Recently, dimer antennas based on colloidal silver NPs have been self-assembled using the DNA origami approach. This structures exhibit a broadband fluorescence enhancement throughout the visible spectral range[158]. DNA origami based dimer antennas were also employed for surface-enhanced Raman spectroscopy (SERS) applications (Figure 4e). Initial experiments addressed an undefined number of molecules[159–161]. Single molecule SERS (Figure 4f) resolution was later attained through a drastic reduction of the interparticle gap. Two different approaches were followed, a silver layer was grown onto the AuNPs[162] (Figure 4g) or shrinking of the DNA origami structure was thermally induced, reaching a gap size of a few nanometers[163]. DNA origami structures have been also recently employed to study the influence of metal NPs on FRET processes[164] (Figure 4h). These studies, which further demonstrate the DNA origami capabilities to self-assemble a pair of fluorophores and a metal NP in a precise geometry, showed that the energy transfer rate between the fluorophores can be moderately enhanced. Additionally, energy transfer along a plasmonic waveguide composed of five metal NPs bound to a DNA origami structure has also been demonstrated[104]. In this work, the energy transfer along 50 nm could be reversibly switched by changing the position of the center particle (Figure 4i).

Not only the interaction of fluorophores and optical antennas consisting of metal NPs has been studied in terms of fluorescence enhancement, SERS and FRET efficiency, but also the first steps towards the analysis of the effect of optical antennas on the emission properties of fluorophores were taken[165]. Through a combination of DNA nanotechnology, plasmonics and super-resolution microscopy, the quantitative study of the emission coupling of single molecules to optical nano-antennas revealed that it can lead to miss-localizations in far-field images (Figure 4j).

Finally, another type of hybrid nanostructure includes the combination of DNA origami structures with the top-down lithographic photonic structures. The first efforts toward the fabrication of these type of hybrid nanostructures included the use of DNA origami structures as sizing units to increase the single occupancy of zero-mode waveguides (also termed nanoapartures)[166] (Figure 4k). This developments could in principle improve the performance of real-time DNA sequencing approaches[167]. Recently, DNA origami structures have been also employed to control the coupling between fluorophores and photonic crystal cavities[168] (Figure 4i). This approach enabled the mapping of the local density of states with subwavelength resolution.

Finally, although the DNA origami arises as the most promising technique to build complex hybrid nanostructures, for some photonic applications, for example in the field of metamaterials, functionality is reached through the combination of several nanostructures arranged in macro-arrays. First steps to creating an array of DNA origami structures were taken with nanoimprint techniques[169] whereas other approaches such as optical printing[170] or STED lithography[171] have not been explored yet.

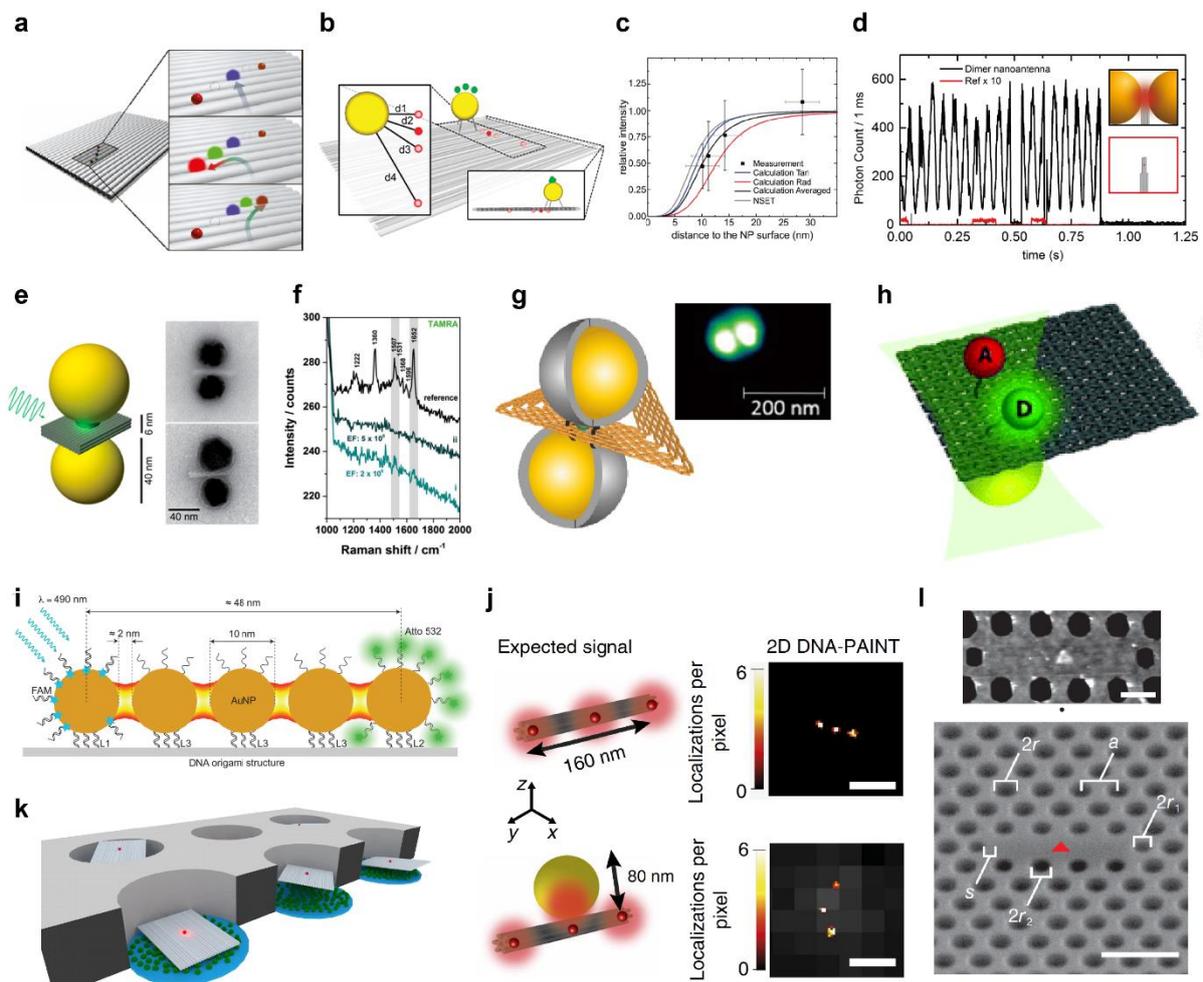

**Figure 4 | DNA origami for hybrid nanophotonic systems. (a)** Directing light through FRET along fluorophores. A single fluorophore (depicted in green) is employed as "jumper"[145]. Sketch of a rectangular DNA origami structure with a single AuNP and fluorophore **(b)** employed to study the distance dependence energy transfer[151] **(c)** Optical antenna dimer for fluorescence enhancement. The incident light polarization is rotated to match the dimer orientation **(d)**. Optical antennas dimer for SERS applications[161] **(e)** Single molecule SERS signal **(f)** obtained with gold dimer antenna covered with silver to further reduce the gap distance[162] **(g)** Sketch

of a DNA origami structure with a single AuNP and a fluorophore pair for plasmonic enhanced FRET studies[164] **(h)** Plasmonic NPs-based optical waveguide for energy transfer over 50 nm[104] **(i)**. **(j)** Study of the emission coupling between single molecules and NPs using super resolution techniques[165]. Combination of DNA origami structures and lithographic plasmonic structures such as zero mode waveguides[166] **(k)** and photonic crystal cavities[168] **(l)**.

# Outlook and future directions

DNA origami-based fabrication of nanophotonic systems has advanced very rapidly in the past decade. It has become a well-established technique for assembly of metal NPs into well-defined clusters including 1D chains, 2D arrays, and 3D lattices [77,81,86,172–177]. The ability to assembling various types of NPs into periodic lattices is very promising for discovery of novel cooperative optical effects. So far, broader applications of the DNA origami technique for plasmonic materials, and materials science in general, have been restricted by the cost of large-scale fabrication. However, very recently biotechnological methods have been successfully developed for mass-production of DNA origami structures, significantly decreasing the cost[178].

Incorporation of dynamic responses into DNA origami-based plasmonic structures affords extra functionalities. In addition, dynamic DNA origami-based plasmonic structures are excellent candidates for generation of smart plasmonic probes for biochemistry and life sciences. One of the main advantages of such probes is the unprecedented level of programmability. Target-probe interactions, transduction mechanisms, and output signals are highly customizable and can be tailored to specific needs. An open challenge is employment of such smart probes also *in vivo* for monitoring cellular processes. Here, utilization of RNA as construction material instead of DNA might provide the solution[179].

In order to realize the full potential of the DNA origami technique and DNA-based imaging approaches in combination with single-molecule fluorescence for the biological and biomedical application, several key challenges have to be solved. One of the main roadblocks moving forward for *in situ* imaging inside cells will be labeling: How can one quantitatively (i.e. 1:1 stoichiometry) and efficiently (ideally 100% target coverage) label proteins inside cell using DNA molecules. DNA origami structures themselves will not be suitable as barcoded labeling probes, due to their extended size, however they can provide a valuable programmable test platform to evaluate novel, orthogonal labeling approaches for proteins such as small molecule binders[180], nanobodies[181], or aptamers[182]. Labeling probes such as the metafluorophores for intensity barcoding discussed above could be adapted to only assemble from small, monomeric units upon detection of a trigger sequence *in situ* inside a cell. In combination, novel labeling and imaging approaches using structural and dynamic DNA nanotechnology could bring Systems Biology to the single cell level, eventually allowing researchers to analyze network-wide interactions of a multitude of biomolecules *in situ* with highest spatial resolution.

The DNA origami technique enabled tremendous progress in the fabrication of hybrid structures for photonics applications. Currently, dimer optical antennas based on metal NPs can reach single molecule SERS sensitivity and outperform lithographic antennas in terms of fluorescence enhancement. These developments render DNA origami based optical antennas into promising devices for diagnostics and DNA sensing applications. Further control on the gap of dimer optical antennas might be exploited for single molecule strong coupling studies[183] among other quantum effects[184]. The coupling between fluorophores and optical antennas might be optimized by controlling the relative orientation. To this end, commercially available doubly linked fluorophores[185] might lead to a much higher control of the fluorophores dipole moment within the DNA origami structure. Along this line, for many applications it would be desirable to replace organic fluorophores with more stable single light sources with improved photophysical properties[185]. Recently QDs[20,186] as well as fluorescent nanodiamonds[41] were successfully incorporated to DNA origami structures. However, nanodiamonds have not been so far combined with optical antennas. Another promising direction is related to fabrication of optical antennas based on high-index dielectric NPs using the DNA origami technique. These particles, of materials such as silicon or germanium among others, attracted considerable attention since they arise as candidates to circumvent one of the main shortcomings of metal NPs which are Joule losses[187].

The DNA origami technique provides flexible platform to meet future nanofabrication needs in nanophotonics. The origami fabrication process uses standard biochemistry lab equipment, e.g, thermoscyclers, centrifuges, etc. Computer-aided design software (caDNAno[28]) and structure predicting tools (CanDo[188]) are intuitive and freely available. All these factors make this technique accessible to anybody with the basic knowledge of DNA. We anticipate that this technique will be more adopted by the general nanophotonic community and will help to complement the existing nanofabrication toolbox. Without a doubt, such adaptations will stimulate a plethora of new exciting research areas and real-life applications.

# Acknowledgements

**Please include research funding is applicable**. Financial support from the Academy of Finland (Grant 308992), the European Research Council (Dynamic Nano), the Volkswagen foundation, and the Sofja Kovalevskaja grant from the Alexander von Humboldt-Foundation; MolMap, Grant agreement number 680241 ), foundation 1, foundation 2, and foundation 3 are gratefully acknowledged.

# References

1. Gwo, S., Chen, H.-Y., Lin, M.-H., Sun, L. & Li, X. Nanomanipulation and controlled self-assembly of metal nanoparticles and nanocrystals for plasmonics. *Chem. Soc. Rev.* **45,** 5672–5716 (2016).
2. Jones, M. R., Osberg, K. D., Macfarlane, R. J., Langille, M. R. & Mirkin, C. A. Templated Techniques for the Synthesis and Assembly of Plasmonic Nanostructures. *Chem. Rev.* **111,** 3736–3827 (2011).
3. Rothemund, P. W. K. Folding DNA to create nanoscale shapes and patterns. *Nature* **440,** 297–302 (2006).
4. Douglas, S. M. *et al.* Self-assembly of DNA into nanoscale three-dimensional shapes. *Nature* **459,** 414–418 (2009).
5. Simmel, F. C. DNA origami – art, science, and engineering. *Front. Life Sci.* **6,** 3–9 (2012).
6. Wang, P., Meyer, T. A., Pan, V., Dutta, P. K. & Ke, Y. The Beauty and Utility of DNA Origami. *Chem* **2,** 359–382 (2017).
7. Tørring, T., Voigt, N. V., Nangreave, J., Yan, H. & Gothelf, K. V. DNA origami: a quantum leap for self-assembly of complex structures. *Chem. Soc. Rev.* **40,** 5636–5646 (2011).
8. Saccà, B. *et al.* Orthogonal Protein Decoration of DNA Origami. *Angew. Chem. Int. Ed.* **49,** 9378–9383 (2010).
9. Sagredo, S. *et al.* Orthogonal Protein Assembly on DNA Nanostructures Using Relaxases. *Angew. Chem. Int. Ed.* **55,** 4348–4352 (2016).
10. Kuzyk, A., Laitinen, K. T. & Törmä, P. DNA origami as a nanoscale template for protein assembly. *Nanotechnology* **20,** 235305 (2009).
11. Mallik, L. *et al.* Electron Microscopic Visualization of Protein Assemblies on Flattened DNA Origami. *ACS Nano* **9,** 7133–7141 (2015).
12. Chandrasekaran, A. R. Programmable DNA scaffolds for spatially-ordered protein assembly. *Nanoscale* **8,** 4436–4446 (2016).
13. Fu, J. *et al.* Assembly of multienzyme complexes on DNA nanostructures. *Nat. Protoc.* **11,** 2243–2273 (2016).
14. Ding, B. *et al.* Gold Nanoparticle Self-Similar Chain Structure Organized by DNA Origami. *J. Am. Chem. Soc.* **132,** 3248–3249 (2010).
15. Hung, A. M. *et al.* Large-area spatially ordered arrays of gold nanoparticles directed by lithographically confined DNA origami. *Nat. Nanotechnol.* **5,** 121–126 (2010).
16. Bui, H. *et al.* Programmable Periodicity of Quantum Dot Arrays with DNA Origami Nanotubes. *Nano Lett.* **10,** 3367–3372 (2010).
17. Wang, R., Nuckolls, C. & Wind, S. J. Assembly of Heterogeneous Functional Nanomaterials on DNA Origami Scaffolds. *Angew. Chem. Int. Ed.* **51,** 11325–11327 (2012).
18. Chao, J., Lin, Y., Liu, H., Wang, L. & Fan, C. DNA-based plasmonic nanostructures. *Mater. Today* **18,** 326–335 (2015).
19. Schreiber, R., Santiago, I., Ardavan, A. & Turberfield, A. J. Ordering Gold Nanoparticles with DNA Origami Nanoflowers. *ACS Nano* **10,** 7303–7306 (2016).
20. Schreiber, R. *et al.* Hierarchical assembly of metal nanoparticles, quantum dots and organic dyes using DNA origami scaffolds. *Nat. Nanotechnol.* **9,** 74–78 (2014).
21. Weller, L. *et al.* Gap-Dependent Coupling of Ag–Au Nanoparticle Heterodimers Using DNA Origami-Based Self-Assembly. *ACS Photonics* **3,** 1589–1595 (2016).

22. Gür, F. N., Schwarz, F. W., Ye, J., Diez, S. & Schmidt, T. L. Toward Self-Assembled Plasmonic Devices: High-Yield Arrangement of Gold Nanoparticles on DNA Origami Templates. *ACS Nano* **10,** 5374–5382 (2016).
23. Dutta, P. K. *et al.* DNA-Directed Artificial Light-Harvesting Antenna. *J. Am. Chem. Soc.* **133,** 11985–11993 (2011).
24. Olejko, L. & Bald, I. FRET efficiency and antenna effect in multi-color DNA origami-based light harvesting systems. *RSC Adv.* **7,** 23924–23934 (2017).
25. Hemmig, E. A. *et al.* Programming Light-Harvesting Efficiency Using DNA Origami. *Nano Lett.* **16,** 2369–2374 (2016).
26. Stein, I. H., Schüller, V., Böhm, P., Tinnefeld, P. & Liedl, T. Single-Molecule FRET Ruler Based on Rigid DNA Origami Blocks. *ChemPhysChem* **12,** 689–695 (2011).
27. Schmied, J. J. *et al.* DNA origami–based standards for quantitative fluorescence microscopy. *Nat. Protoc.* **9,** 1367–1391 (2014).
28. Douglas, S. M. *et al.* Rapid prototyping of 3D DNA-origami shapes with caDNAno. *Nucleic Acids Res.* **37,** 5001–5006 (2009).
29. Douglas, S. M., Chou, J. J. & Shih, W. M. DNA-nanotube-induced alignment of membrane proteins for NMR structure determination. *Proc. Natl. Acad. Sci.* **104,** 6644–6648 (2007).
30. Kick, B., Praetorius, F., Dietz, H. & Weuster-Botz, D. Efficient Production of Single-Stranded Phage DNA as Scaffolds for DNA Origami. *Nano Lett.* **15,** 4672–4676 (2015).
31. Castro, C. E. *et al.* A primer to scaffolded DNA origami. *Nat. Methods* **8,** 221–229 (2011).
32. Bellot, G., McClintock, M. A., Lin, C. & Shih, W. M. Recovery of intact DNA nanostructures after agarose gel-based separation. *Nat. Methods* **8,** 192–194 (2011).
33. Lin, C., Perrault, S. D., Kwak, M., Graf, F. & Shih, W. M. Purification of DNA-origami nanostructures by rate-zonal centrifugation. *Nucleic Acids Res.* **41,** e40 (2013).
34. Shaw, A., Benson, E. & Högberg, B. Purification of Functionalized DNA Origami Nanostructures. *ACS Nano* **9,** 4968–4975 (2015).
35. Stahl, E., Martin, T. G., Praetorius, F. & Dietz, H. Facile and Scalable Preparation of Pure and Dense DNA Origami Solutions. *Angew. Chem. Int. Ed.* **53,** 12735–12740 (2014).
36. Zhang, X., Servos, M. R. & Liu, J. Instantaneous and Quantitative Functionalization of Gold Nanoparticles with Thiolated DNA Using a pH-Assisted and Surfactant-Free Route. *J. Am. Chem. Soc.* **134,** 7266–7269 (2012).
37. Zhang, C. *et al.* A general approach to DNA-programmable atom equivalents. *Nat. Mater.* **12,** 741–746 (2013).
38. Yao, G. *et al.* Clicking DNA to gold nanoparticles: poly-adenine-mediated formation of monovalent DNA-gold nanoparticle conjugates with nearly quantitative yield. *NPG Asia Mater.* **7,** e159 (2015).
39. Kundu, A., Nandi, S. & Nandi, A. K. Nucleic acid based polymer and nanoparticle conjugates: Synthesis, properties and applications. *Prog. Mater. Sci.* **88,** 136–185 (2017).
40. Eskelinen, A.-P. *et al.* Assembly of Single-Walled Carbon Nanotubes on DNA-Origami Templates through Streptavidin–Biotin Interaction. *Small* **7,** 746–750 (2011).
41. Zhang, T. *et al.* DNA-Based Self-Assembly of Fluorescent Nanodiamonds. *J. Am. Chem. Soc.* **137,** 9776–9779 (2015).
42. Voigt, N. V. *et al.* Single-molecule chemical reactions on DNA origami. *Nat. Nanotechnol.* **5,** 200–203 (2010).

43. Kohman, R. E., Cha, S. S., Man, H.-Y. & Han, X. Light-Triggered Release of Bioactive Molecules from DNA Nanostructures. *Nano Lett.* **16,** 2781–2785 (2016).
44. Wagenbauer, K. F. *et al.* How We Make DNA Origami. *ChemBioChem* **18,** 1873–1885 (2017).
45. Linko, V. & Dietz, H. The enabled state of DNA nanotechnology. *Curr. Opin. Biotechnol.* **24,** 555–561 (2013).
46. Hong, F., Zhang, F., Liu, Y. & Yan, H. DNA Origami: Scaffolds for Creating Higher Order Structures. *Chem. Rev.* **117,** 12584–12640 (2017).
47. Linko, V., Ora, A. & Kostiainen, M. A. DNA Nanostructures as Smart Drug-Delivery Vehicles and Molecular Devices. *Trends Biotechnol.* **33,** 586–594 (2015).
48. Bell, N. A. W. & Keyser, U. F. Nanopores formed by DNA origami: A review. *FEBS Lett.* **588,** 3564–3570 (2014).
49. Hernández-Ainsa, S. & Keyser, U. F. DNA origami nanopores: developments, challenges and perspectives. *Nanoscale* **6,** 14121–14132 (2014).
50. Howorka, S. Building membrane nanopores. *Nat. Nanotechnol.* **12,** 619–630 (2017).
51. Rajendran, A., Endo, M. & Sugiyama, H. Single-Molecule Analysis Using DNA Origami. *Angew. Chem. Int. Ed.* **51,** 874–890 (2012).
52. Bald, I. & Keller, A. Molecular processes studied at a single-molecule level using DNA origami nanostructures and atomic force microscopy. *Mol. Basel Switz.* **19,** 13803–13823 (2014).
53. Schmied, J. J. *et al.* Fluorescence and super-resolution standards based on DNA origami. *Nat. Methods* **9,** 1133–1134 (2012).
54. Korpelainen, V., Linko, V., Seppä, J., Lassila, A. & Kostiainen, M. A. DNA origami structures as calibration standards for nanometrology. *Meas. Sci. Technol.* **28,** 34001 (2017).
55. Prodan, E., Radloff, C., Halas, N. J. & Nordlander, P. A Hybridization Model for the Plasmon Response of Complex Nanostructures. *Science* **302,** 419–422 (2003).
56. Halas, N. J., Lal, S., Chang, W.-S., Link, S. & Nordlander, P. Plasmons in Strongly Coupled Metallic Nanostructures. *Chem. Rev.* **111,** 3913–3961 (2011).
57. Hentschel, M. *et al.* Transition from Isolated to Collective Modes in Plasmonic Oligomers. *Nano Lett.* **10,** 2721–2726 (2010).
58. Guerrero-Martínez, A., Grzelczak, M. & Liz-Marzán, L. M. Molecular Thinking for Nanoplasmonic Design. *ACS Nano* **6,** 3655–3662 (2012).
59. Fan, J. A. *et al.* Self-Assembled Plasmonic Nanoparticle Clusters. *Science* **328,** 1135–1138 (2010).
60. Seeman, N. C. DNA in a material world. *Nature* **421,** 427–431 (2003).
61. Mirkin, C. A., Letsinger, R. L., Mucic, R. C. & Storhoff, J. J. A DNA-based method for rationally assembling nanoparticles into macroscopic materials. *Nature* **382,** 607–609 (1996).
62. Alivisatos, A. P. *et al.* Organization of 'nanocrystal molecules' using DNA. *Nature* **382,** 609–611 (1996).
63. Tan, S. J., Campolongo, M. J., Luo, D. & Cheng, W. Building plasmonic nanostructures with DNA. *Nat. Nanotechnol.* **6,** 268–276 (2011).
64. Jones, M. R., Seeman, N. C. & Mirkin, C. A. Programmable materials and the nature of the DNA bond. *Science* **347,** 1260901 (2015).
65. Lan, X. & Wang, Q. DNA-programmed self-assembly of photonic nanoarchitectures. *NPG Asia Mater.* **6,** e97 (2014).
66. Aldaye, F. A., Palmer, A. L. & Sleiman, H. F. Assembling Materials with DNA as the Guide. *Science* **321,** 1795–1799 (2008).

67. Zheng, J. *et al.* Two-Dimensional Nanoparticle Arrays Show the Organizational Power of Robust DNA Motifs. *Nano Lett.* **6,** 1502–1504 (2006).
68. Sharma, J., Chhabra, R., Liu, Y., Ke, Y. & Yan, H. DNA-Templated Self-Assembly of Two-Dimensional and Periodical Gold Nanoparticle Arrays. *Angew. Chem. Int. Ed.* **45,** 730–735 (2006).
69. Le, J. D. *et al.* DNA-Templated Self-Assembly of Metallic Nanocomponent Arrays on a Surface. *Nano Lett.* **4,** 2343–2347 (2004).
70. Park, S. Y. *et al.* DNA-programmable nanoparticle crystallization. *Nature* **451,** 553–556 (2008).
71. Nykypanchuk, D., Maye, M. M., van der Lelie, D. & Gang, O. DNA-guided crystallization of colloidal nanoparticles. *Nature* **451,** 549–552 (2008).
72. Sharma, J. *et al.* Control of Self-Assembly of DNA Tubules Through Integration of Gold Nanoparticles. *Science* **323,** 112–116 (2009).
73. Zhang, C. *et al.* A general approach to DNA-programmable atom equivalents. *Nat. Mater.* **12,** 741–746 (2013).
74. Sharma, J. *et al.* Toward Reliable Gold Nanoparticle Patterning On Self-Assembled DNA Nanoscaffold. *J. Am. Chem. Soc.* **130,** 7820–7821 (2008).
75. Hung, A. M. *et al.* Large-area spatially ordered arrays of gold nanoparticles directed by lithographically confined DNA origami. *Nat. Nanotechnol.* **5,** 121–126 (2010).
76. Pal, S., Deng, Z., Ding, B., Yan, H. & Liu, Y. DNA-Origami-Directed Self-Assembly of Discrete Silver-Nanoparticle Architectures. *Angew. Chem. Int. Ed.* **49,** 2700–2704 (2010).
77. Ding, B. *et al.* Gold Nanoparticle Self-Similar Chain Structure Organized by DNA Origami. *J. Am. Chem. Soc.* **132,** 3248–3249 (2010).
78. Pal, S. *et al.* DNA Directed Self-Assembly of Anisotropic Plasmonic Nanostructures. *J. Am. Chem. Soc.* **133,** 17606–17609 (2011).
79. Dietz, H., Douglas, S. M. & Shih, W. M. Folding DNA into Twisted and Curved Nanoscale Shapes. *Science* **325,** 725–730 (2009).
80. Roller, E.-M. *et al.* DNA-Assembled Nanoparticle Rings Exhibit Electric and Magnetic Resonances at Visible Frequencies. *Nano Lett.* **15,** 1368–1373 (2015).
81. Tian, Y. *et al.* Prescribed Nanoparticle Cluster Architectures and Low-Dimensional Arrays. *Nat. Nanotechnol.* **10,** 637–644 (2015).
82. Urban, M. J. *et al.* Plasmonic Toroidal Metamolecules Assembled by DNA Origami. *J. Am. Chem. Soc.* **138,** 5495–5498 (2016).
83. Liu, Q., Song, C., Wang, Z.-G., Li, N. & Ding, B. Precise organization of metal nanoparticles on DNA origami template. *Methods* **67,** 205–214 (2014).
84. Schreiber, R. *et al.* DNA Origami-Templated Growth of Arbitrarily Shaped Metal Nanoparticles. *Small* **7,** 1795–1799 (2011).
85. Liu, J. *et al.* Metallization of Branched DNA Origami for Nanoelectronic Circuit Fabrication. *ACS Nano* **5,** 2240–2247 (2011).
86. Kuzyk, A. *et al.* DNA-based self-assembly of chiral plasmonic nanostructures with tailored optical response. *Nature* **483,** 311–314 (2012).
87. Pilo-Pais, M., Goldberg, S., Samano, E., LaBean, T. H. & Finkelstein, G. Connecting the Nanodots: Programmable Nanofabrication of Fused Metal Shapes on DNA Templates. *Nano Lett.* **11,** 3489–3492 (2011).
88. Sun, W. *et al.* Casting inorganic structures with DNA molds. *Science* **346,** 1258361 (2014).

89. Helmi, S., Ziegler, C., Kauert, D. J. & Seidel, R. Shape-Controlled Synthesis of Gold Nanostructures Using DNA Origami Molds. *Nano Lett.* **14,** 6693–6698 (2014).
90. Shen, B., Linko, V., Tapio, K., A. Kostiainen, M. & Jussi Toppari, J. Custom-shaped metal nanostructures based on DNA origami silhouettes. *Nanoscale* **7,** 11267–11272 (2015).
91. Shen, X. *et al.* Rolling Up Gold Nanoparticle-Dressed DNA Origami into Three-Dimensional Plasmonic Chiral Nanostructures. *J. Am. Chem. Soc.* **134,** 146–149 (2012).
92. Fan, Z. & Govorov, A. O. Plasmonic Circular Dichroism of Chiral Metal Nanoparticle Assemblies. *Nano Lett.* **10,** 2580–2587 (2010).
93. Shen, X. *et al.* Three-Dimensional Plasmonic Chiral Tetramers Assembled by DNA Origami. *Nano Lett.* **13,** 2128–2133 (2013).
94. Cecconello, A. *et al.* DNA Scaffolds for the Dictated Assembly of Left-/Right-Handed Plasmonic Au NP Helices with Programmed Chiro-Optical Properties. *J. Am. Chem. Soc.* **138,** 9895–9901 (2016).
95. Lan, X. *et al.* Bifacial DNA Origami-Directed Discrete, Three-Dimensional, Anisotropic Plasmonic Nanoarchitectures with Tailored Optical Chirality. *J. Am. Chem. Soc.* **135,** 11441–11444 (2013).
96. Shen, X. *et al.* 3D plasmonic chiral colloids. *Nanoscale* **6,** 2077–2081 (2014).
97. Rao, C. *et al.* Tunable optical activity of plasmonic dimers assembled by DNA origami. *Nanoscale* **7,** 9147–9152 (2015).
98. Lan, X. *et al.* Au Nanorod Helical Superstructures with Designed Chirality. *J. Am. Chem. Soc.* **137,** 457–462 (2015).
99. Liu, H., Shen, X., Wang, Z.-G., Kuzyk, A. & Ding, B. Helical nanostructures based on DNA self-assembly. *Nanoscale* **6,** 9331–9338 (2014).
100. Hentschel, M., Schäferling, M., Duan, X., Giessen, H. & Liu, N. Chiral plasmonics. *Sci. Adv.* **3,** e1602735 (2017).
101. Cecconello, A., Besteiro, L. V., Govorov, A. O. & Willner, I. Chiroplasmonic DNA-based nanostructures. *Nat. Rev. Mater.* **2,** 17039 (2017).
102. Guerrero-Martínez, A., Alonso-Gómez, J. L., Auguié, B., Cid, M. M. & Liz-Marzán, L. M. From individual to collective chirality in metal nanoparticles. *Nano Today* **6,** 381–400 (2011).
103. Klein, W. P. *et al.* Multiscaffold DNA Origami Nanoparticle Waveguides. *Nano Lett.* **13,** 3850–3856 (2013).
104. Vogele, K. *et al.* Self-Assembled Active Plasmonic Waveguide with a Peptide-Based Thermomechanical Switch. *ACS Nano* **10,** 11377–11384 (2016).
105. Roller, E.-M. *et al.* Hotspot-mediated non-dissipative and ultrafast plasmon passage. *Nat. Phys.* **13,** 761 (2017).
106. Zhou, C., Duan, X. & Liu, N. DNA-Nanotechnology-Enabled Chiral Plasmonics: From Static to Dynamic. *Acc. Chem. Res.* (2017). doi:10.1021/acs.accounts.7b00389
107. Zhang, D. Y. & Seelig, G. Dynamic DNA nanotechnology using strand-displacement reactions. *Nat. Chem.* **3,** 103–113 (2011).
108. Krishnan, Y. & Simmel, F. C. Nucleic Acid Based Molecular Devices. *Angew. Chem. Int. Ed.* **50,** 3124–3156 (2011).
109. Schreiber, R. *et al.* Chiral plasmonic DNA nanostructures with switchable circular dichroism. *Nat. Commun.* **4,** (2013).
110. Kuzyk, A. *et al.* Reconfigurable 3D plasmonic metamolecules. *Nat. Mater.* **13,** 862–866 (2014).

111. Kuzyk, A. *et al.* A light-driven three-dimensional plasmonic nanosystem that translates molecular motion into reversible chiroptical function. *Nat. Commun.* **7,** 10591 (2016).
112. Zhan, P. *et al.* Reconfigurable Three-Dimensional Gold Nanorod Plasmonic Nanostructures Organized on DNA Origami Tripod. *ACS Nano* **11,** 1172–1179 (2017).
113. Piantanida, L. *et al.* Plasmon resonance tuning using DNA origami actuation. *Chem. Commun.* **51,** 4789–4792 (2015).
114. Kamiya, Y. & Asanuma, H. Light-Driven DNA Nanomachine with a Photoresponsive Molecular Engine. *Acc. Chem. Res.* **47,** 1663–1672 (2014).
115. Kuzyk, A., Urban, M. J., Idili, A., Ricci, F. & Liu, N. Selective control of reconfigurable chiral plasmonic metamolecules. *Sci. Adv.* **3,** e1602803 (2017).
116. Hu, Y., Cecconello, A., Idili, A., Ricci, F. & Willner, I. Triplex DNA Nanostructures: From Basic Properties to Applications. *Angew. Chem. Int. Ed.* **56,** 15210–15233 (2017).
117. Woo, S. & Rothemund, P. W. K. Programmable molecular recognition based on the geometry of DNA nanostructures. *Nat. Chem.* **3,** 620–627 (2011).
118. Gerling, T., Wagenbauer, K. F., Neuner, A. M. & Dietz, H. Dynamic DNA devices and assemblies formed by shape-complementary, non–base pairing 3D components. *Science* **347,** 1446–1452 (2015).
119. Douglas, S. M., Bachelet, I. & Church, G. M. A Logic-Gated Nanorobot for Targeted Transport of Molecular Payloads. *Science* **335,** 831–834 (2012).
120. Kuzuya, A., Sakai, Y., Yamazaki, T., Xu, Y. & Komiyama, M. Nanomechanical DNA origami 'single-molecule beacons' directly imaged by atomic force microscopy. *Nat. Commun.* **2,** 449 (2011).
121. Walter, H.-K. *et al.* 'DNA Origami Traffic Lights' with a Split Aptamer Sensor for a Bicolor Fluorescence Readout. *Nano Lett.* **17,** 2467–2472 (2017).
122. Gu, H., Chao, J., Xiao, S.-J. & Seeman, N. C. A proximity-based programmable DNA nanoscale assembly line. *Nature* **465,** 202–205 (2010).
123. Zhou, C., Duan, X. & Liu, N. A plasmonic nanorod that walks on DNA origami. *Nat. Commun.* **6,** 8102 (2015).
124. Urban, M. J., Zhou, C., Duan, X. & Liu, N. Optically Resolving the Dynamic Walking of a Plasmonic Walker Couple. *Nano Lett.* **15,** 8392–8396 (2015).
125. Schlichthaerle, T., Strauss, M. T., Schueder, F., Woehrstein, J. B. & Jungmann, R. DNA nanotechnology and fluorescence applications. *Curr. Opin. Biotechnol.* **39,** 41–47 (2016).
126. Hell, S. W. & Wichmann, J. Breaking the diffraction resolution limit by stimulated emission: stimulated-emission-depletion fluorescence microscopy. *Opt. Lett.* **19,** 780–782 (1994).
127. Betzig, E. *et al.* Imaging Intracellular Fluorescent Proteins at Nanometer Resolution. *Science* **313,** 1642–1645 (2006).
128. Rust, M. J., Bates, M. & Zhuang, X. Sub-diffraction-limit imaging by stochastic optical reconstruction microscopy (STORM). *Nat. Methods* **3,** 793–796 (2006).
129. Steinhauer, C., Jungmann, R., Sobey, T. L., Simmel, F. C. & Tinnefeld, P. DNA Origami as a Nanoscopic Ruler for Super-Resolution Microscopy. *Angew. Chem. Int. Ed.* **48,** 8870–8873 (2009).
130. Jungmann, R. *et al.* Single-Molecule Kinetics and Super-Resolution Microscopy by Fluorescence Imaging of Transient Binding on DNA Origami. *Nano Lett.* **10,** 4756–4761 (2010).
131. Derr, N. D. *et al.* Tug-of-War in Motor Protein Ensembles Revealed with a Programmable DNA Origami Scaffold. *Science* **338,** 662–665 (2012).

132. Lin, C. *et al.* Submicrometre geometrically encoded fluorescent barcodes self-assembled from DNA. *Nat. Chem.* **4,** 832–839 (2012).
133. Scheible, M. B., Pardatscher, G., Kuzyk, A. & Simmel, F. C. Single Molecule Characterization of DNA Binding and Strand Displacement Reactions on Lithographic DNA Origami Microarrays. *Nano Lett.* **14,** 1627–1633 (2014).
134. Johnson-Buck, A., Nangreave, J., Jiang, S., Yan, H. & Walter, N. G. Multifactorial Modulation of Binding and Dissociation Kinetics on Two-Dimensional DNA Nanostructures. *Nano Lett.* **13,** 2754–2759 (2013).
135. Johnson-Buck, A. *et al.* Super-Resolution Fingerprinting Detects Chemical Reactions and Idiosyncrasies of Single DNA Pegboards. *Nano Lett.* **13,** 728–733 (2013).
136. Iinuma, R. *et al.* Polyhedra Self-Assembled from DNA Tripods and Characterized with 3D DNA-PAINT. *Science* **344,** 65–69 (2014).
137. Jungmann, R. *et al.* Multiplexed 3D cellular super-resolution imaging with DNA-PAINT and Exchange-PAINT. *Nat. Methods* **11,** 313–318 (2014).
138. Johnson-Buck, A. *et al.* Kinetic fingerprinting to identify and count single nucleic acids. *Nat. Biotechnol.* **33,** 730–732 (2015).
139. Dai, M., Jungmann, R. & Yin, P. Optical imaging of individual biomolecules in densely packed clusters. *Nat. Nanotechnol.* **11,** 798–807 (2016).
140. Jungmann, R. *et al.* Quantitative super-resolution imaging with qPAINT. *Nat. Methods* **13,** 439–442 (2016).
141. Tokunaga, M., Imamoto, N. & Sakata-Sogawa, K. Highly inclined thin illumination enables clear single-molecule imaging in cells. *Nat. Methods* **5,** 159–161 (2008).
142. Schueder, F. *et al.* Universal Super-Resolution Multiplexing by DNA Exchange. *Angew. Chem. Int. Ed.* **56,** 4052–4055 (2017).
143. Schnitzbauer, J., Strauss, M. T., Schlichthaerle, T., Schueder, F. & Jungmann, R. Super-resolution microscopy with DNA-PAINT. *Nat. Protoc.* **12,** 1198–1228 (2017).
144. Woehrstein, J. B. *et al.* Sub–100-nm metafluorophores with digitally tunable optical properties self-assembled from DNA. *Sci. Adv.* **3,** e1602128 (2017).
145. Stein, I. H., Steinhauer, C. & Tinnefeld, P. Single-Molecule Four-Color FRET Visualizes Energy-Transfer Paths on DNA Origami. *J. Am. Chem. Soc.* **133,** 4193–4195 (2011).
146. Bharadwaj, P., Deutsch, B. & Novotny, L. Optical Antennas. *Adv. Opt. Photonics* **1,** 438–483 (2009).
147. Kinkhabwala, A. *et al.* Large single-molecule fluorescence enhancements produced by a bowtie nanoantenna. *Nat. Photonics* **3,** 654–657 (2009).
148. Novotny, L. & van Hulst, N. Antennas for light. *Nat. Photonics* **5,** 83–90 (2011).
149. Koenderink, A. F. Single-Photon Nanoantennas. *ACS Photonics* **4,** 710–722 (2017).
150. Curto, A. G. *et al.* Unidirectional Emission of a Quantum Dot Coupled to a Nanoantenna. *Science* **329,** 930–933 (2010).
151. Acuna, G. P. *et al.* Distance Dependence of Single-Fluorophore Quenching by Gold Nanoparticles Studied on DNA Origami. *ACS Nano* **6,** 3189–3195 (2012).
152. Möller, F. M., Holzmeister, P., Sen, T., Acuna, G. P. & Tinnefeld, P. Angular modulation of single-molecule fluorescence by gold nanoparticles on DNA origami templates. *Nanophotonics* **2,** 167–172 (2013).
153. Pellegrotti, J. V. *et al.* Controlled Reduction of Photobleaching in DNA Origami–Gold Nanoparticle Hybrids. *Nano Lett.* **14,** 2831–2836 (2014).

154. Holzmeister, P. *et al.* Quantum yield and excitation rate of single molecules close to metallic nanostructures. *Nat. Commun.* **5,** 5356 (2014).
155. Acuna, G. P. *et al.* Fluorescence Enhancement at Docking Sites of DNA-Directed Self-Assembled Nanoantennas. *Science* **338,** 506–510 (2012).
156. Vietz, C., Lalkens, B., Acuna, G. P. & Tinnefeld, P. Functionalizing large nanoparticles for small gaps in dimer nanoantennas. *New J. Phys.* **18,** 45012 (2016).
157. Puchkova, A. *et al.* DNA Origami Nanoantennas with over 5000-fold Fluorescence Enhancement and Single-Molecule Detection at 25 µM. *Nano Lett.* **15,** 8354–8359 (2015).
158. Vietz, C., Kaminska, I., Sanz Paz, M., Tinnefeld, P. & Acuna, G. P. Broadband Fluorescence Enhancement with Self-Assembled Silver Nanoparticle Optical Antennas. *ACS Nano* **11,** 4969–4975 (2017).
159. Thacker, V. V. *et al.* DNA origami based assembly of gold nanoparticle dimers for surface-enhanced Raman scattering. *Nat. Commun.* **5,** 3448 (2014).
160. Prinz, J. *et al.* DNA Origami Substrates for Highly Sensitive Surface-Enhanced Raman Scattering. *J. Phys. Chem. Lett.* **4,** 4140–4145 (2013).
161. Kühler, P. *et al.* Plasmonic DNA-Origami Nanoantennas for Surface-Enhanced Raman Spectroscopy. *Nano Lett.* **14,** 2914–2919 (2014).
162. Prinz, J., Heck, C., Ellerik, L., Merk, V. & Bald, I. DNA origami based Au–Ag-core–shell nanoparticle dimers with single-molecule SERS sensitivity. *Nanoscale* **8,** 5612–5620 (2016).
163. Simoncelli, S. *et al.* Quantitative Single-Molecule Surface-Enhanced Raman Scattering by Optothermal Tuning of DNA Origami-Assembled Plasmonic Nanoantennas. *ACS Nano* **10,** 9809–9815 (2016).
164. Aissaoui, N. *et al.* FRET enhancement close to gold nanoparticles positioned in DNA origami constructs. *Nanoscale* **9,** 673–683 (2017).
165. Raab, M., Vietz, C., Stefani, F. D., Acuna, G. P. & Tinnefeld, P. Shifting molecular localization by plasmonic coupling in a single-molecule mirage. *Nat. Commun.* **8,** 13966 (2017).
166. Pibiri, E., Holzmeister, P., Lalkens, B., Acuna, G. P. & Tinnefeld, P. Single-Molecule Positioning in Zeromode Waveguides by DNA Origami Nanoadapters. *Nano Lett.* **14,** 3499–3503 (2014).
167. Eid, J. *et al.* Real-Time DNA Sequencing from Single Polymerase Molecules. *Science* **323,** 133–138 (2009).
168. Gopinath, A., Miyazono, E., Faraon, A. & Rothemund, P. W. K. Engineering and mapping nanocavity emission via precision placement of DNA origami. *Nature* **535,** 401–405 (2016).
169. Gopinath, A. & Rothemund, P. W. K. Optimized Assembly and Covalent Coupling of Single-Molecule DNA Origami Nanoarrays. *ACS Nano* **8,** 12030–12040 (2014).
170. Gargiulo, J., Cerrota, S., Cortés, E., Violi, I. L. & Stefani, F. D. Connecting Metallic Nanoparticles by Optical Printing. *Nano Lett.* **16,** 1224–1229 (2016).
171. Klar, T. A., Wollhofen, R. & Jacak, J. Sub-Abbe resolution: from STED microscopy to STED lithography. *Phys. Scr.* **2014,** 14049 (2014).
172. Liu, W., Li, L., Yang, S., Gao, J. & Wang, R. Self-Assembly of Heterogeneously Shaped Nanoparticles into Plasmonic Metamolecules on DNA Origami. *Chem. – Eur. J.* **23,** 14177–14181 (2017).
173. Tkachenko, A. V., Halverson, J., Gang, O., Liu, W. & Tian, Y. Self-organized architectures from assorted DNA-framed nanoparticles. *Nat. Chem.* **8,** 867 (2016).
174. Ke, Y. *et al.* DNA brick crystals with prescribed depths. *Nat. Chem.* **6,** 994 (2014).

175. Li, H. *et al.* Lattice engineering through nanoparticle–DNA frameworks. *Nat. Mater.* **15,** 654 (2016).
176. Liu, W. *et al.* Diamond family of nanoparticle superlattices. *Science* **351,** 582–586 (2016).
177. Zhang, T. *et al.* 3D DNA origami crystals. *ArXiv170606965 Cond-Mat* (2017).
178. Praetorius, F. *et al.* Biotechnological mass production of DNA origami. *Nature* **552,** 84 (2017).
179. Geary, C., Rothemund, P. W. K. & Andersen, E. S. A single-stranded architecture for cotranscriptional folding of RNA nanostructures. *Science* **345,** 799–804 (2014).
180. Lukinavičius, G. *et al.* A near-infrared fluorophore for live-cell super-resolution microscopy of cellular proteins. *Nat. Chem.* **5,** 132–139 (2013).
181. Ries, J., Kaplan, C., Platonova, E., Eghlidi, H. & Ewers, H. A simple, versatile method for GFP-based super-resolution microscopy via nanobodies. *Nat. Methods* **9,** 582–584 (2012).
182. Opazo, F. *et al.* Aptamers as potential tools for super-resolution microscopy. *Nat. Methods* **9,** 938–939 (2012).
183. Chikkaraddy, R. *et al.* Single-molecule strong coupling at room temperature in plasmonic nanocavities. *Nature* **535,** 127–130 (2016).
184. Zhu, W. *et al.* Quantum mechanical effects in plasmonic structures with subnanometre gaps. *Nat. Commun.* **7,** 11495 (2016).
185. Stennett, E. M. S., Ma, N., van der Vaart, A. & Levitus, M. Photophysical and Dynamical Properties of Doubly Linked Cy3–DNA Constructs. *J. Phys. Chem. B* **118,** 152–163 (2014).
186. Du, K. *et al.* Quantum dot-DNA origami binding: a single particle, 3D, real-time tracking study. *Chem. Commun.* **49,** 907–909 (2013).
187. Baranov, D. G. *et al.* All-dielectric nanophotonics: the quest for better materials and fabrication techniques. *Optica* **4,** 814–825 (2017).
188. Kim, D.-N., Kilchherr, F., Dietz, H. & Bathe, M. Quantitative prediction of 3D solution shape and flexibility of nucleic acid nanostructures. *Nucleic Acids Res.* **40,** 2862–2868 (2012).